\documentclass[prd,aps,showpacs,showkeys]{revtex4-1}
\usepackage{bm,amsfonts,latexsym,amsmath,amssymb,amsbsy,graphicx}

\newcommand{\bb}{\begin{eqnarray}}
\newcommand{\ee}{\end{eqnarray}}
\newcommand{\ba}{\begin{align}}
\newcommand{\ea}{\end{align}}

\begin{document}

\title{\bf  Quantum states of a neutral massive fermion
with an anomalous magnetic moment in an external electric field}
\author{V.R. Khalilov}\email{khalilov@phys.msu.ru}
\affiliation{Faculty of Physics, M.V. Lomonosov Moscow State University, 119991,
Moscow, Russia}

\begin{abstract}
The planar non-relativistic quantum dynamics of a neutral massive fermion
with an anomalous magnetic moment (AMM) in the presence of the electric field of
infinitely long and thin thread with a charge density
distributed uniformly along it (an Aharonov--Casher field) is examined.
The relevant Hamiltonian is singular and requires
additional specification of a one-parameter self-adjoint
extension, which can be given in terms of physically acceptable (self-adjoint)
boundary conditions. We find all possible self-adjoint Hamiltonians
with an Aharonov--Casher field (ACF) by constructing
the corresponding Hilbert space of square-integrable functions,
 including the $r = 0$ region, for all their Hamiltoniahs.
We determine the most relevant physical quantities,
such as energy spectrum and wave functions and discuss
their correspondence with those obtained by the physical regularization
procedure. We show that  energy levels of bound states
are simple poles of the scattering amplitude.
Expressions for the scattering amplitude and cross section depending
on the fermion spin are reexamined.
\end{abstract}



\maketitle


\section{Introduction}

 The Aharonov--Bohm (AB) effect \cite{1} is a quantum physical phenomenon
demonstrating, in particular, the importance of potentials in quantum
mechanics.  Considering  an electron travels in a region
with a non-vanishing  gauge vector potential (since it produces
the magnetic flux in a thin solenoid) but where the magnetic field vanishes,
the electron wave function may develop a quantum (geometric) phase.
 The AB vector potential can produce observable
effects because the relative (gauge invariant) phase of the
electron wave function, correlates with a non-vanishing  gauge
vector potential in the region with zero magnetic field  \cite{khu}.

An interesting and important corollary to the Aharonov--Bohm geometric
phase is a phase acquired by the wave function of a neutral massive fermion
with a magnetic moment when it propagates in an
electric field of a uniformly charged long conducting thread aligned perpendicularly
to the plane of fermion motion. The fermion transport is affected by the phase
acquired with the fermion wave function and the resulting phase difference
leads to a spin-field dependent effects in scattering
(the Aharonov-–Casher effect \cite{ahc}).

The planar quantum dynamics of fermions in the AB potential and ACF  are governed
by  singular Hamiltonians that require the supplementary definition in order
for them to be treated as  self-adjoint quantum-mechanical operators.
Self-adjoint  Hamiltonians are not unique but each of them
can be specified a real "self-adjoint extension" parameter
by additional (self-adjoint) boundary conditions. To put it more exactly,
a domain, including the singular $r = 0$ region, in the corresponding Hilbert space of square-integrable functions must be indicated for each self-adjoint Hamiltonian.
Different choices of self-adjoint extension parameter values  lead
to inequivalent physical cases \cite{phg} so a physical interpretation
of self-adjoint extensions is a purely physical problem and each extension
can  be understood through an appropriate physical
regularization \cite{gtv1}.

We note that self-adjoint Hamiltonians with the AB potential and ACF
have been considered to show the presence of fermion bound
states in \cite{phg,gtv1,kh1,vrkh,safb,khepj1,sil}. Self-adjoint Schr\"odinger
and Dirac Hamiltonians  with singular potentials
(such as the one-dimensional Calogero potential, the Coulomb potential,
a superposition of the Aharonov-–Bohm field and the so-called magnetic-solenoid field and
the potentials localized at the origin, in particular, delta-like
potentials) were considered in many works (see \cite{gtv1} and references there). Also, it should be
noted that the quantum dynamics of a neutral fermion in the cosmic string space-time
have been analyzed  by self-adjoint extension method in Refs. \cite{erbm,facfes}.

In given paper we find all the quantum-mechanical states (wave functions)
for all possible self-adjoint Hamiltonians with the ACF and discuss
their correspondence with those obtained by physical regularization,
which are applied in the convenient quantum mechanics.
These problems for the quantum system under investigation
have not been discussed before in the literature.
 We address the non-relativistic case and find wave functions and energy levels for bound states.
Energy levels for the bound states are derived as  poles of the scattering amplitude.
The scattering problem of spin-polarized neutral fermions in the ACF is briefly
considered. Advanced treatment of the above problems is based on exact solutions
of the self-adjoint Hamiltonians with the ACF, which enables one to examine
these problems the most comprehensive  and complete.

We shall adopt the units where $c=\hbar=1$.

\section{Self-adjoint non-relativistic radial Dirac--Pauli Hamiltonians for
a neutral fermion with AMM in the presence of electric  field}

The Dirac--Pauli equation for a neutral fermion of the mass $m$ with AMM $M$ in
an electric field can be written
in the form of the Schr\"odinger equation as follows
\bb i\frac{\partial\Psi}{\partial t}= H_{DP}\Psi. \label{schr}\ee Here \bb H_{DP}=
\bm{\alpha}\cdot{\bf P}+iM\bm{\gamma}\cdot{\bf E}+\beta m
\label{reham}\ee is the Hamiltonian, ${\bf P}=-i\bm{\nabla}$ is the canonical
momentum operator, $\Psi$ is a bispinor, $\gamma^{\mu}=(\gamma^0,\bm{\gamma}), \bm{\alpha}$ are the Dirac matrices  ${\bf E}$ is the electric field strength.

 Introducing the function \bb
\Psi=\Psi_ne^{-imt} \label{nonfun}\ee and representing $\Psi_n$ in
the form
 \bb \Psi_n = \left(
\begin{array}{c}
\psi\\
\chi
\end{array}\right),
\label{spinor}
 \ee
where $\phi$ and $\chi$ are spinors, we obtain an
equation, which governs the non-relativistic dynamics of  a neutral fermion with the AMM $M$
 in an electric field in the form
\bb i\frac{\partial\psi}{\partial t}=\frac{({\bf
P}-{\bf E}\times {\bf M})^2-M^2{\bf E}^2+M\bm{\nabla}\cdot{\bf E}
}{2m}\psi,\label{fineq}\ee where ${\bf
M}=M\bm{\sigma}$, $\bm{\sigma}$ are the Pauli
 matrices and the term $\bm{\nabla}\cdot{\bf E}$ is equal to
 $4\pi$  times the electric field charge density.

We shall suppose in what follows that the electric field configuration
has the cylindrical symmetry and the system
dynamics is planar. The planar dynamics is accessed by requiring that
the momentum projection $p_z = 0$ together with the imposition of the electric field should
not have a third direction. Then, a natural assumption is that the relevant quantum
 mechanical system is invariant  along the symmetry $z$ axis and the quantum system
moves in the $xy$ plane.
 The electric field generated by infinitely long straight
thin (a zero radius) thread with a charge density $a/2$ distributed uniformly
along the $z$-axis (the ACF) is known to be
\bb E_x=\frac{ax}{r^2},\quad E_y=\frac{ay}{r^2}, \quad
E_z=0,\quad E_r=\frac{a}{r}, \quad E_{\varphi}=0, \label{0one}
 \ee
where  $r=\sqrt{x^2+y^2}, \quad \varphi=\arctan(y/x)$ are polar coordinates.
The Schr\"odinger equation for  a neutral fermion with AMM
 in an electric field (\ref{0one}) takes the form
\bb i\frac{\partial\psi}{\partial t}=\frac{P^2_x+P^2_y +2M\sigma_3(E_xP_y-E_yP_x)+M^2(E_x^2+E_y^2)+ M\bm{\nabla}\cdot{\bf E}
}{2m}\psi,\label{fineq1}\ee
The radial component of the (macroscopic) electric field is determined by the mean surface
charge density as $\bm{\nabla}\cdot{\bf
E}=4\pi \rho$, and the expression
$\rho=a\delta(r)/4\pi r$,
therefore, well approximates $\rho$.
We seek the solutions of (\ref{fineq}) in the ACF  in the polar coordinates in the form
\bb
 \psi(t, r, \varphi) = \frac{\exp(-iEt+ik\varphi)}{\sqrt{2\pi r}} F_E(r), \label{three}
\ee
where $E$ is the particle energy, $k$ is an integer, and $F_E(r)$ is a doublet.

The Hamiltonian of a neutral
fermion in the ACF contains the matrix $\sigma_3$. Therefore, the $\sigma_3$ matrix
commutes with this Hamiltonian and doublet $F_E(r)$ satisfies equation
$\sigma_3F_{E,s}(r)=sF_{E,s}(r)$, where two eigenvalues $s=\pm 1$
correspond two spin projection on the $z$ axis.
Let us denote the upper and lower components of doublet as follows $F_{E,s=1}(r)\equiv F_{E,1}(r)$ and $F_{E,s=-1}(r)\equiv F_{E,2}(r)$; then the scalar product of doublets $F_E(r)$ and $G_E(r)$ is determined by formula \bb
(F_{E}(r),G_{E}(r))=\int (\bar F_{E,1}(r)G_{E,1}(r)+\bar F_{E,2}(r)G_{E,2}(r))dr.
\label{scpr}
\ee
Here $\bar F$ is the complex conjugate function.
Any doublet can be written as
\bb
\left(\begin{array}{c}
AF_{E,1}(r)\\
BF_{E,2}(r)
\end{array}\right)
=AF_{E,1}(r)
\left(\begin{array}{c}
1\\
0
\end{array}\right)+
BF_{E,2}(r)\left(
\begin{array}{c}
0\\
1
\end{array}\right),
\label{doubl}
\ee
where  $A, B$ are  complex constants.

The radial Dirac-Pauli equation for  a neutral
fermion with AMM in the ACF in the  non-relativistic approximation for two components of doublet $F_{E,s}(r)$ can be written in the form
\bb
h\left(\begin{array}{c}
F_{E,1}(r)\\
F_{E,2}(r)
\end{array}\right)
=2mE
\left(\begin{array}{c}
F_{E,1}(r)\\
F_{E,2}(r)
\end{array}\right),
\label{e13}
\ee
where
\bb
h= h_0+Ma\frac{\delta(r)}{r}
\label{2ham}
\ee
and
\bb
h_0=-\frac{d^2}{dr^2} + \frac{k^2+(Ma)^2-1/4}{r^2} +\frac{2kMa}{r^2}\sigma_3.
\label{sym0}
\ee
It is seen from Eqs. (\ref{doubl}) and (\ref{sym0}) that
the upper and lower components of doublet $F_{E,s=1}(r)$ and $F_{E,s=-1}(r)$, in fact,
can be analyzed by means of scalar function $F_{E,s}(r)$ depending explicitly on $s$
 (see, e.g., \cite{ll}). We shall mainly treat  with this function in what follows.

The radial Hamiltonian $h_0$ is  singular and  so the supplementary
definition is required in order for it to be treated as
a self-adjoint quantum-mechanical
operator. Therefore,  we need construct all self-adjoint
Hamiltonians associated with the differential expression
in the right-hand side of Eg.(\ref{sym0}), then specify  correct self-adjoint
extensions by means of physical conditions, i.e., in fact, indicate
the Hamiltonian domain in the Hilbert space of square-integrable functions
on the half-line, including the $r = 0$ region.
The radial Hamiltonian $h$ involves a singular potential term ($Ma\delta(r)/r$),
which influences the behavior of wave functions at $r=0$ and
takes into account by asymptotic self-adjoint boundary conditions
at the origin.

Let us consider the differential expression in the right-hand side of Eg. (\ref{sym0}) as an self-adjoint operator $h^0$ in the Hilbert space  $\mathfrak H=\mathfrak L^2(0,\infty)$ of quantum states for any $(k+sMa)^2-1/4$ but without reservations about its domain. Then, let us just define the operator $h^0$ in the Hilbert space $\mathfrak L^2(0,\infty)$ as
$$
h^0{:}\left\{\begin{array}{l}
 D(h^0)=D(0,\infty), \\
 h^0F(r)= h_0 F(r),
\end{array}\right.
$$
where  $D(0,\infty)$ is the standard space of
smooth functions on $(0,\infty)$ with the compact support
$$
D(0,\infty)={F(r): F(r)\in C^{\infty}, {\rm supp} F \subset [c,d], 0<c<d<\infty}.
$$
This allows us to avoid the problems related to $r\to\infty$.
The operator  $h$ is symmetrical if for any $F(r)$ and $G(r)$
\bb
 \int\limits_{0}^{\infty} \bar G(r)h F(r) dr =
 \int\limits_{0}^{\infty}\bar[h G(r)]F(r) dr.  \label{sym}
 \ee
It is evident that $h^0$ is the symmetrical operator.

Let $h$ be the self-adjoint extension of $h^0$ in $\mathfrak L^2(0,\infty)$ and
let us consider the adjoint operator  $h^*$ given by (\ref{sym0}) but
defined as follows
\bb
h^*{:}\left\{\begin{array}{l}
 D(h^*)=\left\{\begin{array}{l}
F(r):\;F(r)\;\mbox{are absolutely continuous in}(0,\infty),\\
F, h_0 F \in{\mathfrak L}^2(0,\infty),
\end{array}\right.\\
h^*F(r)=h_0F(r),
\label{hadj}
\end{array}\right.
\ee
i.e. $D(h^0)\subset D(h^*)$.
A symmetric operator $h$ is self-adjoint, if its domain $D(h)$
coincides with that of its adjoint operator  $D(h^*)\equiv D^*$.

Integrating (\ref{sym}) by parts and checking that for any function $F(r)$ of $D(h^*)$
$\lim\limits_{r\to \infty} F(r)=0$, we write it in the form
\bb
 (\bar G'F-\bar GF')|_{r=0} =0,  \label{bounsym}
\ee
where  a prime denotes the derivative with respect to $r$.
If (\ref{bounsym}) is satisfied for any $F(r)$ of $D^*$
then the operator  $h^*$ is symmetric and, so, self-adjoint.
This means that  the operator $h^0$ is essentially self-adjoint, i.e.,
its unique self-adjoint extension is its closure $h=\bar h^0$,
which coincides with the adjoint operator $h=h^*$.
If (\ref{bounsym}) is not satisfied then the self-adjoint operator
$h=h^{\dagger}$ can be found as the narrowing  of  $h^*$ on the so-called
maximum domain $D(h)\subset D^*$.
Thus, any $F(r)$ of $D(h)\subset D^*$ must satisfy the self-adjoint boundary condition
\bb
(\bar F'F-\bar FF')|_{r=0} =0. \label{bounsym1}
\ee

Since signs of $a$ and $M$ are fixed it is enough to consider the only
case $Ma>0$. Let us rewrite $Ma$ as follows
\bb
Ma=[Ma]+\mu\equiv n+\mu, \label{intfra}\ee where
$n=0, 1, 2, \ldots$ denotes the largest integer $\le Ma$, and $1>\mu\ge 0$.
It is convenient to change indexing as follows $l\to k+sn$.
Because any solution $F_E(r)$ of Eq. (\ref{e13})  must satisfy asymptotic self-adjoint
boundary condition (\ref{bounsym1}), one can separate out three regions
of the values of $(l+s\mu)^2$.
In first region $(l+s\mu)^2\ge 1$, or  $l+s\mu\ge 1$ and $l+s\mu\le -1$. Since  $-\infty<l<\infty$ and $1>\mu\ge 0$ it easily to determine that all values of $l$  for $\mu>0$ are allowed and
$l=0$, $s=\pm 1$ for $\mu=0$  are not allowed. One can show that for such $l$, the initial symmetric operator $h^0$  is essentially self-adjoint, and its unique self-adjoint extension is  the adjoint operator $h_1=h^*$; its domain  $D(h^*)$ is
\bb
 D(h^*)=\left\{F_E(r):F_E(r), F'_E(r)\mbox{ are absolutely continuous in} (0,\infty),\right. \phantom{mmmmmmm} \nonumber \\ \left. F_E(r), h_0 F_E(r) \in{\mathfrak L}^2(0,\infty), \quad h_1F_E(r)=h_0F_E(r)\right\}.\phantom{mmmmmmmmmm}
\label{1reg}
\ee
The generalized eigenfunctions $F_E(r)$ of the radial Hamiltonian are
\bb
F_E(r)=\sqrt{r}J_{\nu}(pr),
\label{fun1}
\ee
 where $J_{\nu}(pr)$  is the Bessel function
of the order $\nu=|l+s\mu|$ and $p=\sqrt{2mE}$. The energy spectrum is continuous  $E>0$.

In second region $(l+s\mu)^2<1$; it is seen that no values $l$ are allowed for $\mu=0$ and two values $l=0,\mp 1$ for $s=\pm 1$ are allowed for $\mu>0$. Then, for each $l=0,\mp 1$ there exists one-parameter $U(1)$-family of self-adjoint Hamiltonians $h_{\xi}$ parameterized by the real parameter $-\infty\ge\xi\le\infty$ (or $0\ge\theta\le 2\pi$, $\xi=\tan(\theta/2)$); the values
$\xi=\pm \infty$, or $\theta=0,2\pi$ are equivalent. These Hamiltonians are specified by the asymptotic  self-adjoint boundary conditions at the origin with the domain  $D_{\xi}$
\bb
h_\xi{:}\left\{\begin{array}{l}
D_\xi=\left\{\begin{array}{l}
F_E: F_E, F'_E \mbox{are absolutely continuous in}(0,\infty), h_\xi F \in{\mathfrak L}^2(0,\infty),\\
F_E(r)=A[(mr)^{1/2+\gamma_l}+\xi(mr)^{1/2-\gamma_l}]+O(r^3/2), r\to 0,-\infty<\xi<\infty,\\
F_E(r)=A(mr)^{1/2-\gamma_l}+O(r^{1/2}),r\to 0, \xi=\infty
\end{array}\right.\\
 h_\xi F_E(r)= h_0F_E(r),
\end{array}\right.
\label{reg2}\ee
where $A$ is a complex constant and $\gamma_l=||l|-\mu|$.

The energy spectrum of the radial self-adjoint Hamiltonian $h_{\xi}$ is continuous ($E\ge 0$) for any $\xi$ of $-\infty<\xi<\infty$ and the generalized eigenfunctions are
\bb
F_E(r)=C\sqrt{r}\left[J_{\gamma_l}(pr) + \xi\frac{\Gamma(1-\gamma_l)}{\Gamma(1+\gamma_l)}
\left(\frac{-E}{2m}\right)^{\gamma_l}J_{-\gamma_l}(pr)\right],
\label{fun2}
\ee
where $C$ is a constant and $\Gamma(x)$ is the Euler gamma function of argument $x$.

For $-\infty<\xi<0$, there exist  bound fermion states (see, also \cite{khepj1}). In order
for a quantum system to have a bound state, its energy must be negative, and, therefore,
discrete levels with $E<0$  have to exist in addition to continuous part of the energy spectrum.
As was known (see, for example, \cite{ll}) discrete energy levels are simple poles of the scattering
amplitude in the complex plane of $E$ in  the first (physical) sheet of the Riemann surface ${\rm Re}\sqrt{-E}>0$; these poles are the roots of equation
$B_l(E)=0$, where $B_l(E)$ is the coefficient before the ingoing wave in Eq. (\ref{fun2}) at $r\to \infty$:
\bb
B_l(E)=1+\xi\frac{\Gamma(1-\gamma_l)}{\Gamma(1+\gamma_l)}
\left(\sqrt{\frac{-E}{2m}}\right)^{2\gamma_l}.
\label{ingc}
\ee
 After simple calculations, we find the  bound-state energy in the explicit form
\bb
E_l^-=-2m\left(-\xi \frac{\Gamma(1-\gamma_l)}
{\Gamma(1+\gamma_l)}\right)^{-1/\gamma_l}.
\label{enerAC}
\ee
The normalized eigenfunction of bound state $F_l^-$  is
$$
F_l^-(r) =\sqrt{\frac{-2mE_l^-r\sin(\pi \gamma_l)}{\pi \gamma_l}}K_{\gamma_l}(\sqrt{-2mE_l^-}r),
$$
where $K_{\gamma_l}(\sqrt{-2mE_l^-}r)$ is the so-called McDonald function.
Separating Eq. (\ref{enerAC}) as follows
\bb
E_0^-=-2m\left(-\xi \frac{\Gamma(1-\mu)}
{\Gamma(1+\mu)}\right)^{-1/\mu},\quad l=0, s=\pm 1, \gamma_0=\mu, 0<\mu<1,
\label{ener0}
\ee

\bb
E_{\pm 1}^-=-2m\left(-\xi \frac{\Gamma(\mu)}
{\Gamma(2-\mu)}\right)^{-1/(1-\mu)},\quad l=\pm 1, s=\mp 1, \gamma_{\pm 1}=1-\mu.
\label{ener1}
\ee
we see that $E_0^-(\mu)=E_{\pm 1}^-(\mu')$, where $\mu'=1-\mu, 0<\mu'<1$.

Special case $l+s\mu=0$ can be of some interest. In this region,
no values $l$ are allowed for $\mu>0$ and the only value  $l=0 (k=-sn)$ is allowed for $\mu=0$.
For $l=0$ there exists one-parameter $U(1)$-family of
self-adjoint Hamiltonians $h_{\xi}$ parameterized by the real parameter
$-\infty\ge\xi\ge\infty$.
These Hamiltonians are specified by the asymptotic self-adjoint boundary conditions at the origin      with the domain  $D_{\xi}$
\bb
h_\xi{:}\left\{\begin{array}{l}
D_\xi=\left\{\begin{array}{l}
F_E(r), F'_E(r) \mbox {are absolutely continuous in}(0,\infty); F_E(r), h_{\xi}F_E\in{\mathfrak L}^2(0,\infty), \\
F_E(r)=C\sqrt{r}[\ln(mr)+\xi]+ O(r^{3/2}\ln(mr)), r\to 0,\quad -\infty<\xi<+\infty,\\
F_E(r)=C(r)^{1/2}, r\to 0,\quad \xi=\infty
\end{array}\right.\\
 h_\xi F_E(r)= h_0F_E(r),
\end{array}\right.
\label{reg3}\ee
where $C$ is a constant.

One can show  that for $-\infty\ge\xi\ge\infty$ the energy
spectrum is continuous and nonnegative  as well as for  $-\infty<\xi<0$
there  exists one negative level ($E_0$) in addition to the continuous part
of the spectrum
\bb
E_0=-4me^{2(\xi-{\cal C})},
\label{enerAC0}
\ee
where ${\cal C}=0.57721$ is the Euler constant  \cite{GR}.
The generalized eigenfunctions of the continuous spectrum
are the linear combination  of the Bessel ($J_0(pr)$) and Neumann ($N_0(pr)$) functions.
The normalized wave function of bound state  is $\sqrt{-2mE_0r}K_0(\sqrt{-2mE_0}r)$.

\section{Physical regularization procedure}

It should be noted that a choice of a
self-adjoint Hamiltonian requires additional physical arguments \cite{gtv1}.
We emphasize that the radial Hamiltonian $h$ contains a singular potential ($Ma\delta(r)/r$).
 Such a  potential affects  the behavior of wave functions at the origin
but it not grasped by an initial (symmetric) radial Hamiltonian, whose domain
includes functions vanishing at the origin. Mathematically,  such a potential
is grasped by  constructing  self-adjoint extensions of Hamiltonian that
are parameterized  by asymptotic self-adjoint boundary conditions
at the origin so, physically, the nonzero extension parameter
can be treated as a manifestation of additional singular ($\sim \delta(r)/r$) potentials \cite{gtv1}.
For each $\xi$, we find a possible domain for a self-adjoint $h_\xi$ and
different choices $\xi$ lead to inequivalent physical cases (see, also, \cite{phg} and
\cite{asp1,fsa1,as0,sa1}).

 In order to see a correspondence for the model studied
with the physical situation, we solve the problem
using the physical regularization procedure, which is applied in the convenient
quantum mechanics.  We consider a model with the electric field configuration,
which is preferable for the regularization of the $\delta$-function. For this,
the singular electric field (\ref{0one}) is replaced by
\bb
E_z, E_{\varphi}=0, r\ge 0; E_r=0, r<R, E_r=a/r, r>R, mR\ll 1
\label{repl}
\ee
 and the singular two-dimensional
potential $Ma\delta(r)/r$ is replaced  by a regularized one-dimensional potential
$Ma\delta(r-R)/R$ (see, also \cite{khho}).  Such a model implies that the electric field is generated
by an infinitely long straight thin thread (for instance, a conductor) with a surface
charge density distributed uniformly about it along the $z$-axis.
We emphasize that the functional structures
of these configurations are  different and we shall use a well defined
one-dimensional  $Ma\delta(r-R)/R$ term, which can be taken account of by using continuity
conditions at $r=R$.  It is helpful to note that such a term has no $\delta(r)$-function
contribution at the origin (see \cite{crh,safb}).
In these field configuration,  all the solutions of the radial Hamiltonian $h^0$ can
be chosen satisfying the standard boundary condition $F_E(r)=0$. Obviously,
in the region $r<R$, they are the regular functions  $F_E(r)=\sqrt{r}J_{|l|}(pr)$.
Next, we must match the radial solutions and their derivatives in the region $r<R$
with those in the region $r>R$ at $r=R$ taking account of the $Ma\delta(r-R)/R$
potential as
\bb
F_E(R-\delta)= F_E(R+\delta), \quad RF'_E(r)|_{R-\delta}^{R+\delta}=MaF_E(R), \quad \delta\to 0. \label{conmat}
\ee
In the region $r>R$, the eigenfunctions
\bb
F_E(r)=N\sqrt{r}J_{\nu}(pr), \nu\ge 1;
F_E(r)=N_{\pm}\sqrt{r}J_{\pm\gamma_l}(pr), 0<\gamma_l<1
\label{funphys}
\ee
whose coefficients determined by the continuity relations (\ref{conmat})
describe the scattering states, and the eigenfunctions
$F_l^-(r) =N_0\sqrt{r}K_{\gamma_l}(pr), 0<\gamma_l<1$  describe
bound fermion states. We note that the limit $R\to 0$ is to be taken at the calculations of coefficients for wave functions.

Eliminating the normalization constants from Eq. (\ref{conmat})
we obtain a transcendental equation implicitly determining the energy spectrum of
the fermion in the form
\bb
\frac{[\sqrt{X}K_{\gamma_l}(X)]'}{K_{\gamma_l}(X)}-\frac{[\sqrt{X}J_{|l|}(X)]'}{J_{|l|}(X)}=\frac{Ma}{\sqrt{X}},
\label{enbound}
\ee
where $X=\sqrt{2m|E^-|}R$ and now a prime denotes the derivative with respect to $X$.
Taking into account that the regularization parameter $R\ll 1/m$ we can use the following expansions
for the Bessel and McDonald functions
\bb
J_0(z)=1-\frac{z^2}{2}, \quad J_{\nu}(z)=\frac{z^{\nu}}{2^{\nu}\Gamma(1+\nu)}, \quad K_{\nu}(z)=-\frac{\pi}{2\sin(\pi\nu)}
\left[\frac{z^{\nu}}{2^{\nu}\Gamma(1+\nu)}-\frac{z^{-\nu}}{2^{-\nu}\Gamma(1-\nu)}\right].
\label{funcR}
\ee
So, we substitute (\ref{funcR}) into (\ref{enbound}) to calculate the
left-hand side of Eq. (\ref{enbound}).  After these manipulations, we find
the relation:
\bb
\frac{X^{\gamma_l}(|l|+Ma-\gamma_l)}{2^{\gamma_l}\Gamma(1+\gamma_l)}=
\frac{X^{-\gamma_l}(|l|+Ma+\gamma_l)}{2^{-\gamma_l}\Gamma(1-\gamma_l)}.
\label{rel1}
\ee
Solving this equation for $E^-$, we find the following expression for the energy spectrum
of bound states
\bb
E^-=-\frac{2}{mR^2}\left[\frac{(|l|+Ma-\gamma_l)\Gamma(1-\gamma_l)}
{(|l|+Ma+\gamma_l)\Gamma(1+\gamma_l)}\right]^{-1/\gamma_l}, l=0, 0<\gamma_0<1/2; l=\pm 1,
s=\mp 1, 1/2<\gamma_l<1.
\label{energp}
\ee
To be sure, the expression in the square brackets in Eq. (\ref{energp}) must be positive.

Besides, neutral massive fermions can be bound by the field  under discussion,
if the potential, which is the sum of the term $U(r)$ in Eq. (\ref{sym0}) and
a  regularized one-dimensional $Ma\delta(r-R)/R$ potential, is  attractive.
Writing $U(r)$ as $U(r)=(\gamma_l^2-1/4)/r^2, r>R; U(r)=(l^2-1/4)/r^2, r<R$,
we see that  $\gamma_l$ has to be in intervals $0<\gamma_0<1/2$ for $l=0$
and $1/2<\gamma_l<1$ for $|l|=1$ as well as the coupling constant $Ma$ must be negative. Thus,
there exists the only negative level $E^-$ (\ref{ener1}) with $l=0, s=\pm 1$.
Also, these physical arguments are valid regarding Eqs. (\ref{ener0}) and (\ref{ener1}):
only the energy level Eq. (\ref{ener0}) doubly degenerated in spin projection can exist
in the AC background with regularized (at origin) effective potentials.

It should be emphasized that the expressions for energy levels of bound states
(\ref{ener0}) and (\ref{ener1}) depend explicitly upon the self-adjoint extension
parameter $\xi$ and the expression (\ref{energp}) contains the regularization parameter $R$.
Moreover, it follows from parameter of Eq.(\ref{energp})
that the right-hand side of Eq. (\ref{energp})
diverges in the limit $R\to 0$ that is at removing the regularization. Such a behavior of
the energy of bound states is caused by the fact that we treat with the Hamiltonian (\ref{2ham})
in which the term $Ma\delta(r)/r$ is replaced  by $Ma\delta(r-R)/R$. It is essential that
the coupling constant $Ma\equiv |\mu|$ is dimensionless and, therefore, the Hamiltonian
does not contain an initial parameter of the dimension of mass but nevertheless, a bound state
can emerge. So,  the coupling constant must depend on $R$ so that the energy of bound state would remain finite as $R\to 0$ (the renormalization of coupling constant). Thus, the cutoff parameter $\Lambda=1/R$, which (has the dimension of mass and) tends to infinity, transmutes in arbitrary
energy of bound state $E^-$. We believe this is a non-relativistic analog of the phenomenon of dimensional transmutation that occurs in massless relativistic field theories (see, \cite{coulwei}).

\section{The scattering problem}

Now we  discuss the scattering of neutral fermions with AMM  in the considered field.
The particle wave functions for the scattering problem
are constructed by the eigenfunctions given Eqs. (\ref{fun1}) and (\ref{fun2}) (in a self-adjoint
approach and the functions (\ref{funphys}) (in a physical regularization approach).
Since, the fermion wave functions are composed by regular and irregular (but square
integrable) functions we make some remarks. If the singular two-dimensional
potential $Ma\delta(r)/r$ is attractive the singular solutions (more concentrated at the origin
than the regular ones) are to exist, which is quite reasonable from the physical
standpoint. Besides, insisting on regularity of all functions at the origin forces one to reject irregular solutions, which, in some cases, can entail a loss
of completeness in the angular basis.


At first we consider the case when the spatial part of doublet (\ref{three}) $\psi(r,\;\varphi)$  is the eigenfunction of operator $\sigma_3$ with the eigenvalues $s=\pm 1$. Then,
corresponding doublets can be represented in the form
\bb
\left(\begin{array}{c}
\psi_1(r,\;\varphi)\\
\psi_1(r,\;\varphi)
\end{array}\right)
=\sum_{k=-\infty, k\neq -sn}^{\infty}\frac{N_k}{2}
J_{\nu}(pr)e^{ik\varphi}\left(
\begin{array}{c}
1+s\\
1-s
\end{array}\right)+
\frac12
 \left(
\begin{array}{c}
\cos(\theta/2)J_{\mu}(pr)e^{-i\pi\mu/2}(1+s)\\
\sin(\theta/2)J_{-\mu}(pr)e^{i\pi\mu/2}(1-s)
\end{array}\right)e^{-i|n|\varphi}.
\label{expgen}
\ee
where  $N_k=e^{-i\nu\pi/2}e^{-ik\pi/2}, \nu=|k+sn+s\mu|, 1>\mu>0$, $\xi=\tan(\theta/2)$. Expression (\ref{expgen}) is the most general representation of the spatial wave function
of a neutral fermion with AMM for the scattering problem in the AC background.
We emphasize this representation is constructed of solutions (\ref{fun1}) and (\ref{fun2}), which
were found by the self-adjoint extension method.

Now we shall consider the scattering of neutral fermions by electric field (\ref{repl}) and attractive potential $Ma\delta(r-R)/R$. In which case we can essentially simplify the problem
without violating a generality of its consideration. Applying the continuity relations (\ref{conmat})  we present the upper ($\psi_1$), lower ($\psi_2$) component of spatial wave function respectively in the following form
\bb
\left(\begin{array}{c}
\psi_1(r,\;\varphi)\\
\psi_1(r,\;\varphi)
\end{array}\right)
=\sum_{k=-\infty, k\neq -sn}^{\infty}\frac{N_k}{2} J_{\nu}(pr)e^{ik\varphi}\left(
\begin{array}{c}
1+s\\
1-s
\end{array}\right)+
\frac12 \left(
\begin{array}{c}
J_{\mu}(pr)e^{-i\pi\mu/2}(1+s)\\
J_{-\mu}(pr)e^{i\pi\mu/2}(1-s)
\end{array}\right)e^{-in\varphi}.
\label{expsim}
\ee
It is important that the wave function (\ref{expsim}) has been obtained in the limit $R\to 0$
that is as a result of removing the regularization.
It is easily seen that Eq. (\ref{expsim}) corresponds to the values of self-adjoint
extension parameter $\theta=0$ for $s=1$ and $\theta=\pi$ for $s=-1$.

The scattering amplitude is defined in the conventional
manner. We consider the fermion scattering problem assuming that the fermion moves in the positive direction along the $x$ axis before the scattering, i.e., $e^{ipx}$ is an incident wave. In this case, $\varphi$ is the scattering angle. Then, the fermion wave function as $r\rightarrow\infty$ can be written as a superposition of an incident plane wave and a (scattered)
outgoing cylindrical wave
\bb
\psi_p(r,\varphi){\longrightarrow} e^{ipx}+\frac{f(\varphi)}{\sqrt{r}}e^{i(pr-\pi/4)},
\label{assympt}
\ee where $f(\varphi)$ is the scattering amplitude.

The incident plane wave is represented as
$$
e^{ipx}\rightarrow\frac{1}{\sqrt{2\pi{pr}}}    \sum^\infty_{l=-\infty}e^{il\varphi}\left(e^{i(pr-\pi/4)}+(-1)^le^{-i(pr-\pi/4)}\right).
$$
Then, using the asymptotic expansion of Bessel functions at large values of argument
in the form $J_{\pm \nu}(z)= \sqrt{2/\pi z}\cos(z\mp \pi\nu/2-\pi/4)$ the next formula can be easily derived
$$
\psi(r,\varphi)-e^{ipx}=\frac{f_s(\varphi)}{\sqrt{r}}e^{i(pr-\pi/4)},
$$
in which case the scattering amplitude is given by
\bb
f_s(\varphi)=-\frac{s}{\sqrt{2\pi p}}\frac{e^{is(|n|-1/2)\varphi+i|n|\pi}
\sin(\pi\mu)}{\sin(\varphi/2)}u_s,\quad u_s=\frac12\left(
\begin{array}{c}
1+s\\
1-s
\end{array}\right).
\label{asam1}\ee
It is well to note that the asymptotic expansion of scattering solutions
(\ref{assympt}),(\ref{asam1}) does not include the logarithmic phase shift
meanwhile the analogous expansion in the Coulomb field includes the above shift
due to the long-range forces.

Consequently, if the initial-state spin of a fermion moving in the $xy$ plane is oriented
along the $z$ axis, then the fermion scattering cross section in the AC configuration is determined by the formula
\bb
\frac{d\sigma}{d\varphi}=|f(\varphi)|^2=\frac{\sin^2(\pi\mu)}{2\pi p\sin^2(\varphi/2)}.
\label{cros1}\ee
This formula also describes the scattering cross section of non-polarized fermions with an AMM in the
ACF, which was first shown in \cite{ahc}.

Finally, we show that the scattering cross section depends on the initial-state spin of a fermion.
For this, let us consider the case when the spin vector lies in the $xy$ plane.
If, for example, the spatial part of doublet (\ref{three}) $\psi(r,\;\varphi)$  is the eigenfunction of operator $\sigma_1$ with the eigenvalues $s_1=\pm 1$ then the upper ($\psi_1$) and  lower
($\psi_2$) components of spatial wave functions (doublets) can be
determined as
$$
\psi_1(r,\;\varphi)=\sum_{k=-\infty, k\neq -sn}^{\infty}N_k
J_{\nu}(pr)e^{ik\varphi}+[J_{\mu}(pr)e^{i\pi\mu/2}+J_{-\mu}(pr)e^{-i\pi\mu/2}]e^{-i|n|\varphi},
\psi_2(r,\;\varphi)= s_1\psi_1(r,\;\varphi)
$$
and two doublets are
\bb
\left(\begin{array}{c}
\psi_1(r\;\varphi)\\
\psi_1(r\;\varphi)
\end{array}\right), s_1=1;\quad
\left(\begin{array}{c}
\psi_1(r\;\varphi)\\
-\psi_1(r\;\varphi)
\end{array}\right), s_1=-1.
\label{spin1}
\ee
Calculating the scattering amplitudes $f_{\pm}(\varphi)$ for $s_1=\pm 1$ and  cross sections,  we obtain (see, also \cite{khmam})
$$
f_{\pm}(\varphi)=(-1)^{|n|}\frac{\sin(\pi\mu)}{\sqrt{2\pi p}\sin(\varphi/2)}\sin[(|n|-1/2)\varphi]u_{\pm}, \quad u_{\pm}=\frac{1}{\sqrt{2}}\left(
\begin{array}{c}
1\\
\pm 1
\end{array}\right)
$$
and
$$
\frac{d\sigma}{d\varphi}= \frac{\sin^2(\pi\mu)}{2\pi p\sin^2(\varphi/2)}\sin^2[(|n|-1/2)\varphi].
$$


\section{Resume}

We have analyzed the planar quantum dynamics of a neutral massive fermion
with an anomalous magnetic moment in an Aharonov--Casher field.
For this  all the quantum states for all possible self-adjoint
Hamiltonians with the ACF are constructed by applying physically acceptable
(self-adjoint) boundary conditions at origin.  Also, we have found
quantum states of a neutral fermion with AMM by physical regularization procedure
for a model with regularized (at the origin) potentials.
 The problems of bound states and scattering of a neutral massive fermion
with the AMM in the ACF are considered in  detail, with taking into
consideration the fermion spin  and  discussed in terms
of the physics of the problems. In particular, by physical arguments
we show that only one energy level  can exist in the AC background
with regularized effective potentials.

It should be emphasized that
the existence of bound state can affect the scattering process.
Indeed, the bound state can modify the scattering states due the energy
of bound states fixes the scale of energies of the quantum system as a whole.
The total cross-section of non-relativistic particles scattered off a two-dimensional $-\lambda\delta{\bf r}$ potential, in fact, does not change due to
the above modification (see, \cite{khu}). This result
is also correct for the planar scattering  of neutral non-relativistic fermions
with AMM in an Aharonov--Casher field.

There remains one more question.  If the energy $\epsilon$ of scattered fermion, being
a small quantity ($\epsilon R\ll 1$), is close up the energy of bound state,
then the scattering section can significantly increase; this is shown for
scattering of slow particles in an attractive potential $U(r)$ localized
in the region of radius $R$ when there exists a bound $l=0$ - state with the negative
energy small compared with potential $U(r)$ \cite{ll}.
An analysis of the effect of bound state on the planar scattering
of fermions in potentials of an Aharonov--Bohm kind will be deferred to a future
work.

{\bf Acknowledgment}

The author thanks Prof. I.V. Tyutin for valuable comments.

\end{document}